\documentstyle[epsf,11pt]{article}

\makeatletter
\def\eqaligntwo{\stepcounter{equation}\let\@currentlabel=\theequation
\let\\=\@eqncr
$$%%\displ@@
\tabskip\@centering \halign to \displaywidth\bgroup
  \global\@eqcnt\m@ne\hfil
  $\@lign\displaystyle{##}$\tabskip\z@skip&\global\@eqcnt\z@
  $\@lign\displaystyle{{}##}$\hfil\qquad&\global\@eqcnt\@ne
  \hfil$\@lign\displaystyle{##}$&\global\@eqcnt\tw@
  $\@lign\displaystyle{{}##}$\hfil\tabskip\@centering&
  \llap{\@lign##}\tabskip\z@skip\crcr}

\def\endeqaligntwo{\@@eqncr\egroup
      \global\advance\c@equation\m@ne$$\global\@ignoretrue}

\@namedef{eqaligntwo*}{%\@defeqnswfalse
           \eqaligntwo}
\@namedef{endeqaligntwo*}{\endeqaligntwo}
\makeatother

\def\bfp{\hbox{\boldmath $p$}}
\def\bfk{\hbox{\boldmath $k$}}
\def\bfr{\hbox{\boldmath $r$}}
\def\bfA{\hbox{\boldmath $A$}}
\def\Re{\mathop{\rm Re}\nolimits}
\def\Im{\mathop{\rm Im}\nolimits}

\def \cases#1{
           \left \{\,\vcenter {\normalbaselines  \ialign
           {$##\hfil $&\quad {##}\hfil \crcr #1\crcr }}\right .
           }

\begin{document}

\nonfrenchspacing
\flushbottom
\thispagestyle{empty}

\title{\Large\bf  %%% for preprint only
NON-PERTURBATIVE RESULTS FOR HIGH-${T}$ QCD\\[.5cm]
{~}}
\author{R.~JACKIW\thanks{This work is supported in part by funds
provided by the U.S.
Department of Energy (D.O.E.) under cooperative agreement
\#DF-FC02-94ER40818.}\\[1ex]
{\small\em Center for Theoretical Physics, Department of Physics
           and Laboratory for Nuclear Science } \\
{\small\em  Massachusetts Institute of Technology } \\
{\small\em                        Cambridge, MA ~02139~~U.S.A.}}
\maketitle
\thispagestyle{empty}
\bigskip
\begin{center}
To be published in: \\[1ex]
The Proceedings of \\
 {\it VI Mexican School of Particles and Fields \\
Villahermosa, Mexico, October 1994\/} \\
\bigskip\bigskip
MIT-CTP \#2413 \qquad\qquad January 1995
\end{center}
\vspace{1.7cm}
\setlength{\baselineskip}{2.6ex}

These days, as high energy particle colliders become unavailable for testing
speculative theoretical ideas, physicists are looking to other environments
that may provide extreme conditions where theory confronts physical reality.
One such circumstance may arise at high temperature $T$, which perhaps can be
attained in heavy ion collisions or in astrophysical settings.  It is natural
therefore to examine the high-temperature behavior of the standard model, and
here I shall report on recent progress in constructing the high-$T$ limit
of~QCD.

In studying a field theory at finite temperature, the simplest
approach is the so-called imaginary-time formalism.  We continue time
to the imaginary interval $[0,1/iT]$ and consider bosonic (fermionic)
fields to be periodic (anti-periodic) on that interval.  Perturbative
calculations are performed by the usual Feynman rules as at zero
temperature, except that in the conjugate energy-momentum,
Fourier-transformed space, the energy variable $p^0$ (conjugate to the
periodic time variable) becomes discrete --- it is $2\pi n T$, ($n$
integer) for bosons.  From this one immediately sees that at high
temperature --- in the limiting case, at infinite temperature --- the
time direction disappears, because the temporal interval shrinks to
zero.  Only zero-energy processes survive, since ``non-vanishing
energy'' necessarily means high energy owing to the discreteness of
the energy variable $p^0 \sim 2\pi n T$, and therefore all modes with
$n \neq 0$ decouple at large $T$.  In this way a Euclidean
three-dimensional field theory becomes effective at high temperatures
and describes essentially static processes\@.\cite{ref1}

Let me repeat in greater detail. Finite-$T$, imaginary-time
perturbation theory makes use of conventional diagrammatic analysis in
``momentum'' space, with modified ``energy'' variables, as indicated
above. Specifically a spinless boson propagator is
%%
%%% this is a special command from equations.sty
\begin{eqaligntwo*}
D(p)  &=  \frac{i}{p_0^2 - {\bfp}^2 - m^2}
           &   p_0   &=    i\pi (2n) T    \\[0.5ex]
\noalign{\hbox  {while a spin-$\frac{1}{2}$ fermion propagator reads}}
S(p)  &=   \frac{i}{\gamma^0 p_0 - {\hbox{\boldmath $\gamma \cdot p$}} - m}
           &   p_0   &=    i\pi (2n+1) T
\end{eqaligntwo*}
The zero-temperature integration measure $\int \frac{d^4 p}{(2
\pi)^4}$ becomes replaced by\break $iT \sum^{\infty}_{n=-\infty} \int
\frac{d^3 p}{(2 \pi)^3}$.  Thus it is seen that Bose exchange between
two $O(g)$ vertices contributes $iT \sum^{\infty}_{n=-\infty} \int
\frac{d^3 p}{(2 \pi)^3} g \frac{i}{-4 \pi^2 n^2 T^2
-{\bfp}^2-m^2} g$ where $g$ is the coupling strength. In the
large $T$ limit, all $n \neq 0$ terms (formally) vanish and only the
$n = 0$ term survives. One is left with $\int \frac{d^3 p}{(2 \pi)^3}
g \sqrt{T} \frac{1}{\bfp^2 +m^2} g\sqrt{T}$. This is a Bose
exchange graph in a Euclidian 3-dimensional theory, with effective
coupling $g \sqrt{T}$. Similar reasoning leads to the conclusion that
fermions decouple at large~$T$.

While all this is quick and simple, it may be physically inadequate.
First of all, frequently one is interested in non-static processes in
real time, so complicated analytic continuation from imaginary time
needs to be made before passing to the high-$T$ limit, which in
imaginary time describes only static processes.  Also one may wish to
study amplitudes where the real external energy is neither large nor
zero, even though virtual internal energies are high.

Another reason that the above may be inadequate emerges when we
consider massless fields (such as those that occur in QCD). We have
seen that the $n=0$ mode leaves a propagator that behaves as
$\frac{1}{\bfp^2}$ when mass vanishes, and a phase space of
$d^3 p$. It is well known that this kind of kinematics at low momenta
leads to infrared divergences in perturbation theory even for
off-mass-shell amplitudes --- Green's functions in massless Bosonic
field theories possess infrared divergences in naive perturbation
theory\@.\cite{ref4} Since physical QCD does not suffer from
off-mass-shell infrared divergences, perturbation theory must be
resummed.

Thus the formal arguments for the emergence of a 3-dimensional theory
at high-$T$ need be re-examined for QCD\@. Nevertheless, even if
unreliable, the arguments alert us to the possibility that
3-dimensional field theoretic structures may emerge in the high-$T$
regime. Indeed this occurs, although not in a direct, straightforward
fashion; this will be demonstrated presently.

Here is a graphical argument to the same end discussed above: {\em
viz.}~the need to resum perturbation theory.  Consider a one-loop
amplitude $\Pi_1(p)$,
\begin{eqnarray*}
\Pi_1(p) &\equiv& \int dk ~ I_1 (p,k)   \enspace,  \\[0.5pc]
\noalign{\hbox{given by the graph in the figure.}}
%%
%% they warn us against doing this, p.6 of dvips manual
%%

\Pi_1(p) &=&~
\vcenter{\hbox to 125pt{\vbox to 52.8pt{\vss\special{" fig1a}}\hss}}
\\[0.5pc]
&\equiv&  \qquad\quad  \int dk \,  I_1(p,k) \\[0.5pc]
\noalign{\hbox{Compare this to a two-loop amplitude $\Pi_2(p)$,}\vskip 0.5pc}
\Pi_2(p)  &\equiv&   \int dk \,  I_2 (p,k)  \enspace,  \\[0.5pc]
\noalign{\hbox{in which $\Pi_1$ is an insertion, as in the figure below.}
\vskip 0.5pc}
\Pi_2(p) &=& ~
\vcenter{\hbox to 125pt{\vbox to 52.8pt{\vss\special{" fig1b}}\hss}}
\\[0.5pc]
&\equiv& ~~~~~~~~ \, \int dk ~ I_2 (p,k)
\end{eqnarray*}
Following Pisarski,\cite{ref2}
we estimate the relative importance of $\Pi_2$ to $\Pi_1$
by the ratio of their integrands,
$$
{\Pi_2 \over \Pi_1} \sim {I_2 \over I_1} = g^2 \, {\Pi_1(k) \over k^2} ~,
$$
Here $g$ is the coupling constant, and the $k^2$ in the denominator
reflects the fact that we are considering a massless particle, as in QCD\@.
Clearly the $k^2 \to 0$ limit is relevant
to the question whether the higher order graph can be neglected relative to
the lower order one.
Because one finds that for small $k$ and large
$T$, $\Pi_1(k)$ behaves as $T^2$, the ratio $\Pi_2/\Pi_1$ is $g^2 T^2 / k^2$.
As a result when $k$ is $O(gT)$ or smaller the two-loop amplitude is not
negligible compared to the one-loop amplitude.  Thus graphs with ``soft''
external momenta [$O(gT)$ or smaller] have to be
included as insertions in higher order calculations.

A terminology has arisen: graphs with generic/soft external moment [$O
(g T)\/$ where $g\/$ is small and $T\/$ is large] and large internal
momenta [the internal momenta are integration variables in an
amplitude; when $T\/$ is large they are $O (T)\/$, hence also large]
are called ``hard thermal loops\@.''\cite{ref2,ref3} Much study has
been expended on them and finally a general picture has emerged.
Before presenting general results, let us look at a specific example
--- a 2-point Green's function.

It needs to be appreciated that in the imaginary-time formalism the
correlation functions are unique and definite.  But passage to real
time, requires continuing from the integer-valued ``energy'' to a
continuous variable, and this cannot be performed uniquely.  This
reflects the fact that in real time there exists a variety of
correlation functions: time ordered products, retarded commutators,
advanced commutators, {\em etc}.  Essentially one is seeing the
consequence of the fact that a Euclidean Laplacian possesses a unique
inverse, whereas giving an inverse for the Minkowskian d'Alembertian
requires specifying temporal boundary conditions, and a variety of
answers can be gotten with a variety of boundary conditions.

Thus, when presenting results one needs to specify precisely what one is
\hbox{computing}.

We shall consider a correlation function for two fermionic currents,
in the 1-loop approximation.
\begin{eqnarray*}
\Pi^{\mu \nu} (x,y)
           &=&  -i \left< j^\mu (x) j^\nu (y) \right>   \\
&=&  \int \frac{d^4 k}{(2 \pi)^4}  e^{-i k (x - y)}
           \Pi^{\mu \nu} (k)
\end{eqnarray*}
The QCD result differs from the QED result by a group theoretical
multiplicative factor, so we present high-$T\/$ results only for the
latter, in real-time, and consider the time-ordained product $\Pi^{\mu
\nu}_T\/$ as well as the retarded commutator $\Pi^{\mu \nu}_R\/$.

$\Pi^{\mu \nu}\/$ possesses a real and an imaginary part.  It is found
that at large $T\/$, the real parts of $\Pi^{\mu \nu}_T\/$ and
$\Pi^{\mu \nu}_R\/$ coincide.
$$
- {\Re} \Pi^{\mu \nu} (k) = \frac{T^2}{6} P^{\mu \nu}_2
           + \frac{T^2 k^2}{| {\bfk}^2 |}
           \left[ 1 + \frac{k^0}{2 | {\bfk} |}
           \ln \left| \frac{k^0 - | {\bfk} |}{k^0 + | {\bfk} |} \right| \right]
           \left[ \frac{1}{3} P^{\mu \nu}_1 + \frac{1}{2} P^{\mu \nu}_2 \right]
$$
where the projection operators are
\begin{eqnarray*}
P^{\mu \nu}_1  &=&  g^{\mu \nu} - k^\mu k^\nu / k^2  \\
P^{\mu \nu}_2  &=&
          \cases{
                 0  &  if $\mu\/$ or $\nu = 0$    \cr
                 \delta^{i j} - k^i k^j /  | {\bfk}^2 |  & otherwise \cr
           }  %end of cases
\end{eqnarray*}
For the imaginary part, which is present only for space-like
arguments, different expressions are found.
\begin{eqnarray*}
- {\Im} \Pi^{\mu \nu}_R (k)
           &\equiv&  - \pi \rho^{\mu \nu} (k)  \\
&=&  \frac{\pi k^2}{| {\bfk} |^3} \frac{k^0}{2}
           T^2 \theta (-k^2)
           \left[ \frac{1}{3} P^{\mu \nu}_1 + \frac{1}{2} P^{\mu \nu}_2 \right]
                       \\
- {\Im} \Pi^{\mu \nu}_T (k)
           &=& \frac{\pi k^2}{| {\bfk} |^3}
           \left[ \frac{k^0}{2} T^2 + T^3 \right]
           \theta (-k^2)
           \left[\frac{1}{3} P^{\mu \nu}_1 + \frac{1}{2} P^{\mu \nu}_2 \right]
\end{eqnarray*}

A unified presentation of these formulas is achieved in a dispersive
representation.  For the retarded function this reads
\begin{eqnarray*}
\Pi^{\mu \nu}_R ({k})
           &=&  \Pi^{\mu \nu}_{S U B} (k)
           + \int d k'_0
           \frac{\rho^{\mu \nu} (k'_0, {\bfk})}{k'_0 - k_0 - i \epsilon}  \\
\noalign{\hbox {while the time-ordered expression is}}
\Pi^{\mu \nu}_T (k)
           &=&  \Pi^{\mu \nu}_{S U B} (k)
           + \int d k'_0
           \frac{\rho^{\mu \nu} (k'_0, {\bfk})}{k'_0 - k_0 - i \epsilon}
           + \frac{2 \pi i}{e^{k_0 / T} - 1}
           \rho^{\mu \nu} (k_0, {\bfk})
\end{eqnarray*}
The dispersive expressions may also be used to give the imaginary-time formula.
$$
\Pi^{\mu \nu}_{\rm imaginary \atop time} (k)
           = \Pi^{\mu \nu}_{S U B} (k)
           + \int d k'_0
           \frac{\rho^{\mu \nu} (k'_0, {\bfk})}{k'_0 - 2 \pi i n T}
$$
In all the above formulas, $\Pi^{\mu \nu}_{S U B} \/$ is a real
subtraction term.

Note that a universal statement about high-$T\/$ behavior can be made
only for the absorptive part $\rho^{\mu \nu} :\/$ it is $O (T^2)\/$.
This also characterizes $\Pi^{\mu \nu}_R\/$, but $\Pi^{\mu \nu}_T\/$
possesses an additional $O (T^3)\/$ imaginary part, which is seen to
arise from the additional term in $\Pi^{\mu \nu}_T\/$ involving the
bosonic distribution function $\frac{1}{e^{k_0 / T} - 1}\/$.  Finally,
the $\Pi^{\mu \nu}_{\rm imaginary \atop time}\/$ amplitude has a
temperature behavior determined by its external ``energy'' $= 2 \pi i
n T\/$.  If this is replaced by a fixed $k_0\/$ ($T\/$-independent) or
if only the $n = 0\/$ mode is considered, then one may assign an $O
(T^2)\/$ behavior to this quantity as well.

In conclusion, we assert that the 2-point correlation function behaves
as $O (T^2)\/$, where it is understood that this statement is to be
applied to the retarded amplitude, or to the imaginary time amplitude
with its ``energy'' argument continued away from $2 \pi i n
T\/$\@.\cite{ref9}

Similar analysis has been performed on the higher-point functions and
this work has culminated with the discovery (Braaten, Pisarski,
Frenkel, Taylor)\cite{ref2,ref3,ref5} of a remarkable simplicity in
their structure.  To describe this simplicity, we do not discuss the
individual $n\/$-point functions, but rather their sum multiplied by
powers of the vector potential, {\em viz.}~we consider the generating
functional for single-particle irreducible Green's functions with
gauge field external lines in the hard thermal limit.  (Effectively,
we are dealing with continued imaginary-time amplitudes.)  We call
this quantity $\Gamma_{\rm H T L} (A)\/$ and it is computed in an
$SU(N)$ gauge theory containing $N_F$ fermion species of the
fundamental representation.  $\Gamma_{\rm H T L}\/$ is found (i) to be
proportional to $(N+{1\over2} N_F)$, (ii) to behave as $T^2$ at high
temperature, and (iii) to be gauge invariant.
\def\myfig#1{%%
\vcenter{\hbox to 52.8pt{\vbox to 54.9pt{\vss\special{" #1}}\hss}}}
\begin{eqnarray*}
\myfig{2 0 doit}~+\myfig{3 -90 doit}+\myfig{4 45 doit}
           &+&  \myfig{5 90 doit}
           ~+ \cdots  \\
\noalign{\vskip-1ex}
&&\hskip2ex =  (N + {\textstyle{1\over2}} N_F) \, {g^2 T^2 \over 12\pi} \,
           \Gamma_{\rm HTL} (A) \\
\noalign{\vskip1ex}
\Gamma_{\rm HTL} (U^{-1} \, A \, U + U^{-1} \, dU)  \hskip-4ex
&&\hskip2ex =
           \Gamma_{\rm HTL} (A)
\end{eqnarray*}
(Henceforth $g\/$, the coupling constant, is scaled to unity.)
A further kinematical simplification in
$\Gamma_{\rm HTL}$ has also been established.
To explain this we define two light-like four-vectors
$Q^\mu_{\pm}$
depending on a unit three-vector $\hat{q}$,
pointing in an arbitrary direction.
$$
Q^\mu_\pm  =  {1\over\sqrt{2}} (1,\,\pm \hat{q})
$$
\begin{eqnarray*}
\hat{q} \cdot \hat{q}  = 1  \enspace,  \qquad
           Q^\mu_\pm Q_{\pm \mu}  &=&  0 \enspace,  \qquad
           Q^\mu_\pm Q_{{\mp} \mu}  = 1
\end{eqnarray*}
Coordinates and potentials are projected onto $Q_\pm^\mu$.
$$
x^\pm \equiv x_\mu Q_\pm^\mu  \enspace,  \qquad
           \partial_\pm \equiv Q_\pm^\mu {\partial \over \partial x^\mu}
                      \enspace,  \qquad
           A_\pm \equiv A_\mu Q_\pm^\mu
$$
The additional fact that is now known is that (iv)
after separating an ultralocal contribution from
$\Gamma_{\rm HTL}$, the remainder may be written as an average over
the angles of
$\hat{q}$
of a functional $W$ that  depends only on $A_+$;
also this functional
is non-local only on the
two-dimensional $x^\pm$ plane, and is ultralocal in the remaining directions,
perpendicular to the $x^\pm$ plane.  [``Ultralocal'' means that any potentially
non-local kernel $k(x,y)$ is in fact a $\delta$-function of the difference
$k(x,y)  \propto  \delta (x-y)$.]
$$
\Gamma_{\rm HTL} (A) = 2\pi \int d^4x ~ A^a_0 (x) A^a_0(x) +
\int d\Omega_{\hat{q}} \, W (A_+)
$$
These results are established in perturbation theory, and a perturbative
expansion of $W(A_+)$, {\it i.e.\/}~a power series in $A_+$, exhibits the
above mentioned properties.  A natural question is whether one can sum the
series to obtain an expression for $W(A_+)$.

Important progress on this problem was made when it was observed
(Taylor, Wong)\cite{ref5}
that the gauge-invariance condition
can be imposed infinitesimally, whereupon it leads
to a functional differential equation for $W(A_+)$, which is best presented as
\begin{eqnarray*}
&&{\partial \over \partial x^+} \, {\delta \over \delta A^a_+}
           \left[ W(A_+) + {\textstyle {1\over2}}
                      \int d^4 x ~ A_+^b(x) A_+^b(x) \right]  \\
&& {\qquad\qquad\qquad}  - {\partial \over -\partial x^-}
           \left[ A_+^a \right] + f^{abc} A_+^b
           {\delta \over \delta A_+^c}
           \left[ W(A_+) + {\textstyle {1\over2}}
                      \int d^4 x ~ A_+^d(x) A_+^d(x) \right]
           = 0
\end{eqnarray*}
In other words we seek a quantity, call it
$$
S(A_+) \equiv W(A_+) +{1\over2} \int
d^4 x \, A_+^a (x) A_+^a(x)   \enspace,
$$
which is a functional on a two-dimensional
manifold $\left\{ x^+, x^- \right\}$, depends on a single functional variable
$A_+$, and satisfies
$$
\partial_1 {\delta \over \delta A_1^a} S - \partial_2 A_1^a +
f^{abc} A_1^b {\delta \over \delta A_1^c} S = 0
$$
$$
{\hbox{``1''}} \equiv x^+   \enspace,  \qquad
{\hbox{``2''}} \equiv -x^-   \enspace,  \qquad
A_1^a \equiv A_+^a
$$
Another suggestive version of the above is gotten by defining $A_2^a \equiv
{\delta S \over \delta A_1^a}$.
$$
\partial_1 A_2^a - \partial_2 A_1^a + f^{abc} A_1^b A_2^c = 0
$$
To solve the functional equation and produce an expression for $W(A_+)$, we
now turn to a completely different corner of physics,
and that is Chern-Simons theory at zero temperature.

The Chern-Simons term is a peculiar
gauge theoretic topological
structure that  can be constructed in odd
dimensions, and here we consider it in  3-dimensional space-time.
$$
I_{\rm CS} \propto \int d^3 x \, \epsilon^{\alpha \beta \gamma} \,
{\rm Tr} \,
(\partial_\alpha A_\beta A_\gamma + {\textstyle {2\over3}} A_\alpha A_\beta
A_\gamma)
$$
This object was introduced into physics over a decade ago, and since that time
it has been put to various physical and mathematical uses.
Indeed
one of our originally stated motivations
for studying the Chern-Simons term
was its possible relevance
to high-temperature gauge theory\@.\cite{ref6}
Here following Efraty and Nair,\cite{ref7}
we shall employ the Chern-Simons term
for a determination of the hard
thermal loop generating functional, $\Gamma_{\rm HTL}$.

Since it is the space-time integral of a density, $I_{\rm CS}$ may be viewed
as the action for a quantum field theory
in (2+1)-dimensional space-time,
and the corresponding Lagrangian
would then be given by a two-dimensional, spatial integral
of a Lagrange density.
\begin{eqnarray*}
I_{\rm CS} &\propto&  \int dt ~ L_{\rm CS} \\
L_{\rm CS} &\propto&  \int d^2 x  \,
           \left(A_2^a \skew{4}\dot{A}_1^a + A_0^a F_{12}^a \right)
\end{eqnarray*}
I have separated the temporal index (0) from the two spatial ones (1,2) and
have indicated time differentiation
by an over dot.
$F_{12}^a$ is the non-Abelian field strength, defined on a two-dimensional
plane.
$$
F_{12}^a = \partial_1 A_2^a - \partial_2 A_1^a + f^{abc} A_1^b A_2^c
$$
Examining the Lagrangian, we see that it has the form
$$
L \sim p \dot{q} - \lambda \, H(p,q)
$$
where $A_2^a$ plays the role of $p$, $A_1^a$ that of $q$, $F^a_{12}$
is like a Hamiltonian and $A^a_0$ acts like the
Lagrange multiplier $\lambda$, which forces the Hamiltonian
to vanish; here $A_0^a$ enforces the vanishing of $F_{12}^{\,a}$.
$$
F_{12}^{\,a} = 0
$$
The analogy instructs us how the Chern-Simons
theory should be quantized.

We postulate equal-time commutation relations, like those between
$p$ and~$q$.
$$
\left[ A_1^a ({\bfr}), \, A_2^b ({\bfr}') \right]
           = i \, \delta^{ab} \delta({\bfr} - {\bfr}')
$$
In order to satisfy the condition enforced by the Lagrange multiplier, we
demand that $F_{12}^a$, operating on ``allowed'' states, annihilate them.
$$
F_{12}^a | ~~ \rangle = 0
$$

This equation can be explicitly presented in a Schr\"odinger-like
representation
for
the Chern-Simons quantum field theory, where the state is a functional of
$A_1^a$.  The action of the operators $A_1^a$ and $A_2^a$ is by multiplication
and functional differentiation, respectively.
\begin{eqnarray*}
\phantom{A_0^a} \, | ~~ \rangle &\sim&  \Psi(A_1^a) \\
A_1^a \, | ~~ \rangle &\sim&  A_1^a \, \Psi(A_1^a) \\
A_2^a \, | ~~ \rangle &\sim&  {1\over i} {\delta \over \delta A_1^a}
\, \Psi(A_1^a)
\end{eqnarray*}
This, of course, is just the field theoretic analog of the quantum mechanical
situation where states are functions of $q$, the $q$ operator acts by
multiplication, and the $p$ operator by differentiation.
In the Schr\"odinger representation,
the condition that states be annihilated by $F_{12}^a$
$$
\left( \partial_1 A_2^{a} - \partial_2  A_1^a + f_{abc} A_1^b
A_2^c \right) \, \Big| ~~ \Big\rangle = 0
$$
leads to a functional differential equation.
$$
\left(
\partial_1 {1\over i} {\delta \over \delta A_1^a}
- \partial_2 \, A_1^a
+ f_{abc} A_1^b \, {1\over i} \, {\delta \over \delta A_1^c}
\right)
\Psi(A_1^a) = 0
$$
If we define $S$ by $\Psi = e^{iS}$ we get equivalently
$$
\partial_1 {\delta \over \delta A_1^a} S - \partial_2 A_1^a + f_{abc} A_1^b
{\delta \over \delta A_1^c} S = 0
$$
This equation comprises the entire content of Chern-Simons quantum field
theory.
$S$ is the Chern-Simons eikonal, which gives the exact wave functional owing
to the simple dynamics of the theory.
Also the above eikonal equation is
recognized to be precisely the equation
for the hard thermal loop generating functional, given above.

Let me elaborate on the connection with eikonal-WKB ideas.  Let us
recall that in particle quantum mechanics, when the wave function
$\psi (q)\/$ is written in eikonal form
$$
\psi (q) = e^{i S (q)}
$$
then the WKB approximation to $S (q)\/$ is given by the integral of
the canonical 1-form $p d q\/$
$$
S (q) = \int^q p (q') d q'
$$
where $p (q)\/$, the momentum, is taken to be function of the
coordinate $q\/$, by virtue of satisfying the equation of motion.
\begin{eqnarray*}
\frac{p^2 (q)}{2} + V (q)  &=&  E  \\
p(q)  &=&  \sqrt{2 E - 2 V (q)}
\end{eqnarray*}
Analogously, in the present field theory application, the eikonal $S
(A_1)\/$ may be written as
$$
S (A_1) = \int^{A_1} A^a_2 (A'_1) {\cal D} A'^a_1
$$
where $A^a_2 (A_1)\/$ is functional of $A_1\/$ determined by the
equation of motion
$$
\partial_1 A^a_2 - \partial_2 A^a_1 + f^{a b c} A^b_1 A^b_2 = 0
$$
Since, by construction $\frac{\delta S}{\delta A^a_1} = A^a_2\/$, it
is clear that as a consequence $S\/$ satisfies the required equation.
However, we reiterate that here there is no WKB approximation:
everything is exact.

The gained advantage for thermal physics is that ``acceptable''
Chern-Simons states, {\it i.e.\/}~solutions to the above functional
equations, were constructed long ago,\cite{ref8} and one can now take
over those results to the hard thermal loop problem.  One knows from
the Chern-Simons work that $\Psi$ and $S$ are given by a 2-dimensional
fermionic determinant, {\it i.e.\/}~by the Polyakov-Wiegman
expression.  While these are not described by very explicit formulas,
many properties are understood, and the hope is that one can use these
properties to obtain further information about high-temperature QCD
processes.

For example one can give a very explicit series expansion for
$\Gamma_{\rm HTL}\/$ in terms of powers of~$A\/$
$$
\Gamma_{\rm HTL} = {{1}\over{2!}}
           \int \Gamma^{(2)}_{\rm HTL} AA + {{1}\over{3!}} \int
           \Gamma^{(3)}_{\rm HTL} AAA + \cdots
$$
where the non-local kernels $\Gamma^{(i)}_{\rm HTL}\/$ are known
explicitly.  This power series may be used to systematize the
resummation procedure for pertubative theory.
Here is what one does:  perturbation theory for Green's functions may
be organized with the help of a functional integral, where the
integrand contains (among other factors) $e^{i I_{\rm QCD} (A)}\/$
where $I_{\rm QCD}\/$ is the QCD action.  We now rewrite that as
$$
e^{i \left\{ I_{\rm QCD} (A) + {{m^2}\over{4 \pi}}
           \Gamma_{\rm HTL} (A) - {{m^2}\over{4 \pi}}
           \Gamma_{\rm HTL} (A) \right\} }
$$
where $m =  T \sqrt{{{N + N_F / 2}\over{3}}}\/$.  Obviously nothing
has changed, because we have merely added and subtracted the
hard-thermal-loop generating functional.  Next we introduce a loop
counting parameter $l\/$: in an $l\/$-expansion, different powers of
$l\/$ correspond to different numbers of loops, but at the end $l\/$ is
set to unity.  The resummed action is then taken to be
$$
e^{i I_{\rm resummed}} = e^{i \left\{ {{1}\over{l^2}} \left[ I_{\rm QCD} (l A)
           + {{m^2}\over{4 \pi}} \Gamma_{\rm HTL} (l A) \right]
           - {{m^2}\over{4 \pi}} \Gamma_{\rm HTL} (l A) \right\} }
$$
One readily verifies that an expansion in powers of $l\/$
describes the resummed perturbation theory, free of infrared
divergences.

Even though the closed form for $\Gamma_{\rm HTL}\/$ is not very
explicit, a much more explicit formula can be gotten for its
functional derivative ${{\delta \Gamma_{\rm HTL}}\over{\delta
A^a_\mu}}\/$.  This may be identified with an induced current, which
is then used as a source in the Yang-Mills equation.  Thereby one
obtains a non-Abelian generalization of the Kubo equation, which
governs the response of the hot quark gluon plasma to external
disturbances\@.\cite{ref9}
$$
D_\mu F^{\mu\nu} =
{m^2 \over 2} j^\nu_{\rm induced}
$$
{}From the known properties   of the fermionic determinant --- hard thermal
loop
generating functional --- one can show that $j^\mu_{\rm induced}$ is given by
$$
j^\mu_{\rm induced} = \int {d \Omega_{\hat{q}} \over 4\pi} \, \left\{
Q_+^\mu \left( \vphantom{1\over1} a_-(x) - A_-(x) \right)
+ Q_-^\mu \left( \vphantom{1\over1} a_+ (x) - A_+(x) \right) \right\}
$$
where $a_{\pm}$ are solutions to the equations
\begin{eqnarray*}
\partial_+ a_- - \partial_- A_+ + [A_+, a_-] &=& 0 \\
\partial_+ A_- - \partial_- a_+ + [a_+, A_-] &=&  0
\end{eqnarray*}
Evidently $j^\mu_{\rm induced}$, as determined by the above equations, is a
non-local and non-linear functional of the vector potential $A_\mu$.

There now have appeared several alternative derivations of the Kubo
equation.  Blaizot and Iancu\cite{ref10} have analyzed the
Schwinger-Dyson equations in the hard thermal regime; they truncated
them at the 1-loop level, made further kinematical approximations that
are justified in the hard thermal limit, and they too arrived at the
Kubo equation.  Equivalently the argument may be presented succinctly
in the language of the composite effective action,\cite{ref11} which
is truncated at the 1-loop (semi-classical) level --- two-particle
irreducible graphs are omitted.  The stationarity condition on the
1-loop action is the gauge invariance constraint on $\Gamma_{\rm
HTL}\/$.  Finally, there is one more, entirely different derivation
--- which perhaps is the most interesting because it relies on
classical physics\@.\cite{ref12} We shall give the argument presently,
but first we discuss solutions for the Kubo equation.

To solve the Kubo equation, one must determine $a_{\pm}\/$ for arbitrary
$A_{\pm}\/$, thereby obtaining an expression for the induced current,
as a functional of $A_\pm\/$.  Since the functional is non-local and
non-linear, it does not appear possible to construct it explicitly
in all generality.  However, special cases can be readily handled.

In the Abelian case, everything commutes and linearizes.
One can determine $a_\pm$ in terms of $A_{\pm}$.
$$
a_\pm = {\partial_\pm \over \partial_{\mp}} \, A_{\mp}
$$
Incidentally, this formula exemplifies the kinematical simplicity,
mentioned above, of hard thermal loops:
the nonlocality of $1/\partial_\pm$ lies entirely in the
$\left\{ x^+, x^- \right\}$
plane.   With the above form for $a_\pm$ inserted into the
Kubo equation, the solution can be constructed explicitly.
It coincides with the results obtained by Silin long ago,
on the basis of the Boltzmann-Vlasov equation\@.\cite{ref13}
One sees that the present theory is the non-Abelian
generalization of that physics;  in particular $m$, given above,
is recognized as the Debye screening length,
which remains gauge invariant in the non-Abelian context.

It is especially interesting to emphasize that Silin did not use
quantum field theory in his derivation; rather he employed classical
transport theory.  Nevertheless, his final result coincides with what
here has been developed from a quantal framework.  This raises the
possibility that the non-Abelian Kubo equation can also be derived
classically, and indeed such a derivation has been given, as mentioned
above.

We now pause in our discussion of solutions to the non-Abelian Kubo
equation in order to describe its classical derivation.

Transport theory is formulated in terms of a single-particle
distribution function $f\/$ on phase space.  In the Abelian case,
$f\/$ depends on position $\{ x^\mu \}\/$ and momentum $\{ p^\mu \}\/$
of the particle.  For the non-Abelian theory it is necessary to
take into account the fact that the particle's non-Abelian charge
$\{Q^a \}\/$ also is a dynamical variable: $Q^a\/$ satisfies an evolution
equation (see below) and is an element of phase space.  Therefore, the
non-Abelian distribution function depends on $\{ x^\mu \}\/$,
$\{p^\mu \}\/$ and $\{ Q^a \}\/$, and in the collisionless
approximation obeys the transport equation ${{d}\over{d \tau}} f = 0\/$,
{\it i.e.\/}
$$
{{\partial f}\over{\partial x^\mu}} {{d x^\mu}\over{d \tau}} +
           {{\partial f}\over{\partial p^\mu}} {{d p^\mu}\over{d \tau}} +
           {{\partial f}\over{\partial Q^a}} {{d Q^a}\over{d \tau}} = 0
$$
The derivatives of the phase-space variables are given by the Wong
equations, for a particle with mass $\mu$.
\begin{eqnarray*}
{{d x^\mu}\over{d \tau}}  &=&  {{p^\mu}\over{\mu}}   \\
{{d p^\mu}\over{d \tau}}  &=&  F^{\mu \nu}_a {{d x_\nu}\over{d \tau}} Q^a \\
{{d Q^a}\over{d \tau}}  &=&  - f^{a b c}  {{d x^\mu}\over{d \tau}}
                                 A^b_\mu Q^c  \\
\end{eqnarray*}
In order to close the system we need an equation for $F^{\mu \nu}\/$.
In a microscopic description (with a single particle) one would have
$(D_\mu F^{\mu \nu})^a =
 \int d \tau Q^a {(\tau)} {{p^\nu (\tau)}\over{\mu}}
\delta^4 \big( x - x (\tau) \big)\/$
and consistency would require covariant
conservation of the current; this is ensured
provided $Q^a\/$ satisfies the equation given above.  In our
macroscopic, statistical derivation, the current is given in terms of the
distribution function, so the system of equations closes with
$$
(D_\mu F^{\mu \nu})^a  = \int d p \, d Q \, Q^a p^\nu f (x, p, Q)
$$
The collisionless transport equation, with the equations of
motion inserted, is called the Boltzmann equation.  The closed system
formed by the latter supplemented with the
Yang-Mills equation is known as the non-Abelian Vlasov equations.
To make progress, this highly non-linear set of equations is
approximated by expanding around the equilibrium form for $f\/$,
$$
f^{\raise 1ex \hbox{$\scriptstyle \rm free$}}
           _{\hskip-0.5ex\lower 1.5ex \hbox{$
                      {\scriptstyle \rm boson \hfill}
                      \atop{\scriptstyle \rm fermion}$}}
           \propto
        \left(e^{ {{1}\over{T}} \sqrt{ \bfp^2 + \mu^2} }
                      {\scriptstyle\mp} 1\right)^{-1}
$$
This comprises the Vlasov approximation, and readily leads to the
non-Abel\-ian Kubo equation\@.\cite{ref12}

One may say that the non-Abelian theory is the minimal elaboration of
the Abelian case needed to preserve non-Abelian gauge invariance.  The
fact that classical reasoning can reproduce quantal results is
presumably related to the fact that the quantum theory makes use of
the (resummed) 1-loop approximation, which is frequently reorganized
as an essentially classical effect.  Evidently, the quantum
fluctuations included in the hard thermal loops coincide with thermal
fluctuations.

Returning now to our summary of the solutions to the non-Abelian Kubo
equation that have been obtained thus far, we mention first that the
static problem may be solved completely\@.\cite{ref11} When the {\it
Ansatz\/} is made that the vector potential is time independent,
$A_\pm = A_\pm ({\bfr})\/$, one may solve for $a_\pm\/$ to find $a_\pm
= - A_\pm\/$ and the induced current is explicitly computed as
$$
{{m^2}\over{2}} j^\mu_{\rm induced} =
           {{-m^2 A^0 }\choose{ \bf 0 }}
$$
This exhibits gauge-invariant electric screening with Debye mass
$m\/$.  One may also search for localized static solutions to the Kubo
equation, but one finds only infinite energy solutions, carrying a
point-magnetic monopole singularity at the origin.  Thus there are no
plasma solitons in high-T QCD\@.\cite{ref11} Specifically, one finds
that upon selecting the radially symmetric solution, which decreases
at large distances, there arises a magnetic monopole-like singularity
at the origin.

Much less is known concerning time-dependent solutions.  Blaizot and
Iancu\cite{ref14} have made the {\it Ansatz\/} that the vector
potentials depend only on the combination $x \cdot k\/$, where $k\/$
is an arbitrary 4-vector: $A_\pm = A_\pm (x \cdot k)\/$.  Once again
$a_\pm\/$ can be determined; one finds $a_\pm = {{Q_\pm \cdot
k}\over{Q_{\mp} \cdot k}} A_{\mp}\/$, and the induced current is
computable.  For $k = \left( {{1}\atop{{\bf 0}}} \right)\/$, where
there is no space dependence (only a dependence on time is present)
one finds
$$
{{m^2}\over{2}} j^{\mu}_{\rm induced} =
           {0 \choose {- {{1}\over{3}} m^2 {\bfA} }}
$$
More complicated expressions hold with general $k\/$.  The Kubo
equation can be solved numerically; the resulting profile is a
non-Abelian generalization of a plasma plane wave.

The physics of all these solutions, as well as of other, still undiscovered
ones, remains to be elucidated, and I invite any of you to join in this
interesting task.

\end{document}